\documentclass[sigconf,nonacm]{acmart}

\usepackage[most]{tcolorbox}
\usepackage{hyperref}

\AtBeginDocument{%
  }

\begin{document}

\title{Using psychological theory to ground guidelines for the annotation of misogynistic language}

\author{Artemis Deligianni}
\email{A.Deligianni@ed.ac.uk}
\affiliation{
  \institution{University of Edinburgh}
  \country{United Kingdom}
}

\author{Zachary Horne}
\email{Zachary.Horne@ed.ac.uk}
\affiliation{
  \institution{University of Edinburgh}
  \country{United Kingdom}
}

\author{Leonidas A. A. Doumas}
\email{Alex.Doumas@ed.ac.uk}
\affiliation{
  \institution{University of Edinburgh}
  \country{United Kingdom}
}

\renewcommand{\shortauthors}{Deligianni et al.}

\begin{abstract}
 Detecting misogynistic hate speech is a difficult algorithmic task. The task is made more difficult when decision criteria for what constitutes misogynistic speech are ungrounded in established literatures in psychology and philosophy, both of which have described in great detail the forms explicit and subtle misogynistic attitudes can take. In particular, the literature on algorithmic detection of misogynistic speech often rely on guidelines that are insufficiently robust or inappropriately justified -- they often fail to include various misogynistic phenomena or misrepresent their importance when they do. As a result, current misogyny detection coding schemes and datasets fail to capture the ways women experience misogyny online. This is of pressing importance: misogyny is on the rise both online and offline. Thus, the scientific community needs to have a systematic, theory informed coding scheme of misogyny detection and a corresponding dataset to train and test models of misogyny detection. To this end, we developed (1) a  misogyny annotation guideline scheme informed by theoretical and empirical psychological research, (2) annotated a new dataset achieving substantial inter-rater agreement ($\kappa$ = 0.68) and (3) present a case study using Large Language Models (LLMs) to compare our coding scheme to a self-described "expert" misogyny annotation scheme in the literature. Our findings indicate that our guideline scheme surpasses the other coding scheme in the classification of misogynistic texts across 3 datasets. Additionally, we find that LLMs struggle to replicate our human annotator labels, attributable in large part to how LLMs reflect mainstream views of misogyny. We discuss implications for the use of LLMs for the purposes of misogyny detection.
\end{abstract}


\maketitle

\begin{figure}
    \centering
    \includegraphics[width=0.4\linewidth]{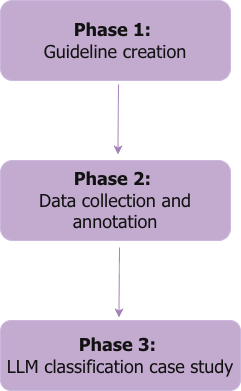}
    \caption{Phases of the methodological workflow.}
    \label{fig:flowchart}
\end{figure}

\section{Introduction}
In 2025, the United Nations warned that the rise in misogynistic content online is driving backlash against feminism and women's rights  \cite{un_2025_manopshere}. Extreme forms of misogyny are becoming mainstream in online spaces. And as scholars long warned the public, online misogyny \textit{is} offline misogyny \cite{Jane_2017, kottig_at_2017}. For example, a UK survey by Hope not Hate found that 45\% of 16-24 year old male respondents had a positive opinion of Andrew Tate, a man who became famous largely because of his misogynistic views shared via major social media platforms like TikTok, Instagram and YouTube \cite{andrew_tate_UK}. Both secondary and primary school teachers report that male students are repeating and enacting misogynistic content they consume online  \cite{over2025understanding}. 

Simultaneously, the content moderation policies of many social media platforms are failing to shield and may even discriminate against the minority groups that they claim to protect \cite{diaz2021double}. Black and transgender creators face disproportionate removals of their content, often including discussions of their marginalised identities \cite{haimson2021disproportionate}. Beyond failing to successfully moderate hate speech, many of the new features developed on social media platforms are being used to engage in new forms of hateful acts. For instance, in 2026, thousands of women (and children) had images of themselves digitally undressed by users prompting X's Grok chatbot \cite{elon_X_grok}. 
Clearly, establishing and enforcing safeguards is necessary to prevent the implementation of malicious requests like these.

Still, a major challenge faces the automated misogyny detection systems that scientists are developing. Hate speech detection literature has an \textit{inconsistency} problem. Several datasets for the detection of sexism, racism and beyond, define analogous concepts inconsistently, resulting in incompatible datasets and in turn leading to divergent behaviours in trained-classifiers \cite{fortuna_survey_2019}. In large part, this inconsistency across the literature points to the larger problem that there is a general lack of theoretical and empirical foundation in our annotation guidelines. \cite{dutta2025divided} reflect on how inconsistency in the definitions of sexism (or misogyny) across papers makes their analytic comparison a challenging task. They highlight the need for interdisciplinarity, particularly emphasising the importance of insights from social sciences in order to tackle misogyny in all its complex and nuanced forms and call for a theoretically grounded exploration of online sexism \cite{dutta2025divided}. This paper's main aim is to meet this call. 

Here, we provide a set of guidelines and an annotated dataset for misogyny detection which incorporate several decades worth of theoretical and empirical work on sexism from psychological science. We present results from a comparative case study of our guidelines and dataset against those guidelines and the dataset presented in \cite{guest_expert_2021}. Specifically, we experimentally test the performance of three Large Language Models (LLMs) at applying both guidelines schemes across the two annotated datasets and analyse their class-specific performance statistically to provide a more nuanced method of comparison. High performance on a given dataset does not necessarily mean a model is robust and more comprehensive evaluation of models is typically required \cite{dutta2025divided}. Therefore, our third contribution involves adapting items from the Ambivalent Sexism Inventory to serve as a small test dataset \cite{glick_ambivalent_1996}. This inventory is one of the well-known theoretical frameworks for understanding the multifaceted nature of sexism \cite{glick_ambivalent_1996}. We use this adapted dataset to test the adherence of two guidelines schemes we consider to the underlying psychological theory of Ambivalent Sexism. Our results indicate that compared to the guideline scheme developed by \cite{guest_expert_2021}, our guidelines consistently improve LLMs' classification of misogynistic items without reducing their performance on non-misogynistic items. Moreover, we find strong evidence to suggest not only that our guidelines improve the LLMs' performance on the dataset we developed from the Ambivalent Sexism Inventory items, but that the guidelines developed by \cite{guest_expert_2021} degrade performance on these items relative to a baseline condition. Still, we find that the addition of our guidelines to the LLMS' prompts does not fully mitigate their inbuilt biases, and the overall performance of these models remains poor. We discuss implications for the use of LLMS in the detection of misogyny.

\subsection{Background}

\subsubsection{Limitations of existing annotation guidelines}
There are a number of frequently cited datasets for misogyny detection in the literature. Some are specifically created for misogyny detection and some are created for hate speech detection more broadly but also include misogyny or sexism in their categories. As noted above, a systematic analysis of their contents revealed a lack of theoretical grounding \cite{dutta2025divided}. Below we introduce some of these datasets and attempt to identify some possibilities for why annotation guidelines for misogyny detection might produce inconsistencies. 

First, some papers apply unsuitable theoretical frameworks to the annotation of misogyny. For example, \cite{talat2016hateful} applied the same coding scheme developed from critical race theory to both the annotation of racism and sexism, without motivating why this approach is sensible for the annotation of sexism. Though it is undeniable that parallels that can be found in the way both racism and sexism are expressed, as both are forms of prejudice \cite{gordon1965nature}, the resultant coding scheme is far too general to train and aid annotators to identify the various specific ways misogyny can be expressed. Lending credence to our point is the fact the paper reports that 85\% of their annotation disagreements occurred for the annotation of sexism, even though texts targetting women were less common \cite{talat2016hateful}. The authors themselves acknowledge this potential shortcoming, as they attribute a large number of their disagreements in annotation to "the difference of opinion in what constitutes sexism" between the annotators \cite{talat2016hateful}. 

Even when attempts are made to ground guideline criteria on the empirical work on online misogyny, the scope of the theoretical work is quite narrow and often focuses on very particular forms of misogyny that do not capture the full breadth of its expression. For example, \cite{anzovino2018automatic} classify misogynistic Twitter posts according to a set of criteria developed from  a single book about cybersexism \cite{poland2016haters}. In this book the author analyses a series of events and research into online sexism to draw parallels between online and offline sexism \cite{poland2016haters}. The resultant coding scheme proposes classifying misogynistic language in terms of its linguistic function: Discrediting, Stereotype and Objectification, Sexual Harassment and Threats of Violence, Dominance, and Derailing \cite{anzovino2018automatic}. It is certainly the case that misogyny can strive to enact these things on women in various ways, and this list is non-exhaustive (for example the work by \cite{manne2017down}). However, the resultant coding scheme presents some significant limitations. The first limitation is that these guidelines define misogyny by the outcome an individual intends to achieve through the language they use and not the content of what they say. Defining misogynistic language by its outcome overemphasises the intention behind a statement and fails to define how misogyny is \textit{ employed} in that statement. In short, annotators need to speculate about the intention of the author, instead of focusing on the what was actually said, and inferring intent from text is often difficult. As a result, the inter-rater agreement for both the binary classification task and the further sub-classification task suffer (moderate and low respectively \cite{anzovino2018automatic}). 

Other papers which attempt to build upon and improve such guideline schemes, inadvertently inherit the limitations of the original guideline schemes. For example, both  \cite{guest_expert_2021,basile2019semeval} base their guidelines on  \cite{anzovino2018automatic}. \cite{guest_expert_2021} then augment and extend the categories of the original coding scheme. The authors explain that these additions to the coding scheme aim to improve on the preceding guidelines by further dividing misogynistic behaviours into different forms of misogynistic abuse \cite{guest_expert_2021}. The authors report overall moderate levels of inter-rater agreement (Fleiss' $\kappa =0.484$, with the highest level of agreement being for misogynistic pejoratives (Fleiss' $\kappa$=0.559) and the lowest agreement for misogynistic personal attacks (Fleiss' $\kappa$=0.145). This low agreement is perhaps unsurprising given the low inter-rater reliability in the work that these guidelines expanded upon. 

\subsection{Why psychological theory?} 
In the misogyny detection literature, misogyny is often defined as a type of language which is intended to harm women in particular ways (e.g., by discrediting, objectifying or sexually harassing them \cite{anzovino2018automatic, guest_expert_2021}). Whilst is is true that misogyny can do all of the above and more, the approach risks defining misogyny by its intended consequences rather than its underlying form. That is, this approach tells annotators that misogyny is identified by the intended outcome (belittlement, harassment, discrediting, etc.) of the language used.This task requires a good deal of potential speculation on the part of the annotator, and risks missing, at the very least, cases of unintentional (but still effective) misogynistic speech.

Beyond the difficulty of interpreting the intentions behind any potential misogynistic text, it is incorrect to assume that every individual enacting misogynistic speech considers themselves a misogynist, let alone intends to be one. For example, qualitative work shows that often people use disclaimers such as "I'm not sexist, but..." before following with a sexist statement, even in discussions as serious as intimate partner violence \cite{harris2012m}. Additionally, \cite{manne2017down} (pp.43-44) has discussed at length how the general naive view tends to pathologise  misogyny, presenting it as a condition or mental state to be diagnosed in the individual actor and not a social system enacted by people more broadly to be criticised and dismantled.  

Ideally, a definition and guideline scheme of misogyny should provide annotators with insights for how specific misogynistic ends are achieved through language. In the psychological literature, misogyny is comprised of a set of different beliefs and biases about women that can be theoretically defined, analysed and studied empirically \cite{zehnter2021belief, glick_ambivalent_1996, jones2023gender}. 
Basing our definition and guidelines on this work might not only provide a more consistent definition of misogyny, but also explain the ways that it is framed, expressed and defended through various forms of argument. Analysing the argument put forth in a text does not require us to make assumptions about the mental state of the individual expressing sexist or misogynistic beliefs in order to judge whether a statement is sexist or not. Under this framework, annotators need only be familiar with common beliefs and biases and, most importantly, the ways that these can be expressed in the text. 

Indeed, previous work has benefited from the use of psychological theory in its development. However, this work is often focused on very specific subcategories or subtypes of sexism. For instance,  \cite{jha2017does}, used Ambivalent Sexism Theory to develop a coding scheme for classifying various subcategories of hostile and benevolent sexism. This work is important, but our goals here are to develop a classification system for detecting misogynistic language more broadly.

\subsubsection{History of Feminism in Psychology}
Feminism has done a great deal for the social advancement of women's rights and has also provided critical insights into the design of scientific research and data analysis (e.g. \cite{d2023data}). Psychology has a long history of incorporating feminist perspectives in its theories and methods \cite{eagly2012feminism,rutherford2015feminism}. Starting in the 1960s, feminists criticised psychology for not only neglecting the study of women and gender more broadly, but for often misrepresenting women in both its theory and empirical research \cite{eagly2012feminism,rutherford2015feminism}. Feminist activism paved the ground for the emergence of feminist psychology, which has given rise to many theories, methods, and topics of study \cite{rutherford2015feminism}. Much of this work now features in mainstream psychology, not just in more specialised gender journals \cite{rutherford2015feminism}. As a consequence, many of the current theoretical perspectives in psychology, while not typically labelled feminist, reflect feminist thought and values\cite{rutherford2015feminism}. As psychology combines feminist thought with scientific rigour, its theories and empirical findings provide a reasonable basis for grounding our misogyny annotation guidelines.

\subsubsection{Online and offline misogyny}
A common approach in the current literature is to focus specifically on misogyny as it manifests online. While it is clear that online misogyny is an important issue, separating misogynistic language into online and offline forms at least implicitly supports a notion that the kind of misogyny that women experience in their everyday lives is fundamentally different from that experienced online. We disagree with this idea, and echo the sentiments of previous works that have argued that the only thing about online misogyny is the medium (the internet) and not the sentiment being expressed towards women \cite{Jane_2017}. Our goal is to provide a set of annotation guidelines that captures the phenomenon is it manifests in language regardless of medium. 

\begin{table*}[!ht]
\caption{Examples for the concepts included in our misogyny annotation guidelines. \textit{Internalised misogyny} is not listed separately– any woman endorsing the concepts listed below would constitute internalised misogyny.}
\centering
\renewcommand{\arraystretch}{1.3}
\begin{tabular}{ll}
\toprule
\textbf{Misogyny Concept}    & \textbf{Example(s)}                                              \\
\midrule
Hostile Sexism      & Women are gold-diggers.                              \\
Benevolent Sexism   & Women need men to protect them.                      \\
Gender Essentialism & Men are naturally smarter than women.                \\
Toxic Masculinity   & Real men don't cry.                                  \\
Gendered Racism     & She is just an angry black woman.                    \\
Post-Feminism       & (1) Men and women are already equal; \\
{ } & (2) Women no longer experience discrimination \\
Backlash            & Feminism has gone too far and now men are oppressed. \\
\bottomrule 
\end{tabular}
    \label{tab:guidelines}
\end{table*}

\section{Guidelines}

Our full guidelines are available online at \href{https://osf.io/3c8kp/overview?view_only=5ba9d2a050d34fc3bc84fdd27ef46e58}{Open Science Framework (OSF)}\footnote{All files can be found under the Files tab; There is no need to create an account or request access to download the guidelines or data.}. Before we detail the concepts included in our guidelines and how they might help improve on the current state of misogyny detection, we want to highlight a terminological issue: psychological theories tend to use the term "sexism" when describing prejudiced beliefs, attitudes and behaviours against women. Therefore, we treat the terms "sexism" and "misogyny" as interchangeable in this work. For further discussion of this issue please see page 4 of the guidelines \footnote{\url{https://osf.io/3c8kp/overview?view_only=5ba9d2a050d34fc3bc84fdd27ef46e58}}.
 
To capture the various ways misogyny is expressed, including more subtle sexism, we leveraged the psychological literature into prejudice, sexism and misogyny. The resulting annotation guidelines include the following well-documented phenomena: ambivalent sexism, gender essentialism, toxic masculinity, gendered forms of racism, post-feminism, backlash and internalized misogyny.

\textbf{Ambivalent sexism} captures the idea that hostile attitudes towards women can be endorsed through benevolently (for the speaker) sexist attitudes for women \citep{glick_ambivalent_1996}. Benevolent sexism presents men's power over women as a positive,, often framing men as providers and protectors of women and praising conformity to traditional gender roles. In contrast, hostile sexism demonizes women and portrays them as villains who are taking advantage of men \cite{hammond_benevolent_2018}. For example, a benevolent sexist statement could be "men should protect women" or "women are great caregivers", whereas hostile statements can range from gendered slurs to negative characterisations such as, "women use men for money". The misogyny detection literature has previously included discussions of the "severity" of misogynist language in a manner that seems to loosely reflect ambivalent sexism. However, framing ambivalent sexism in terms of severity is inaccurate and harmful, as there is nothing necessarily less severe about experiencing benevolent sexism \cite{glick_ambivalent_1996}. 

\textbf{Gender essentialism} defines beliefs that gender differences are inherent and "natural" or biologically determined (e.g., that men are naturally smarter than women) \cite{skewes_beyond_2018}. Gender essentialism posits a variety of innate qualities differentiating men and women that are not supported in the scientific literature, such as skills and interests \cite{skewes_beyond_2018}. Gender essentialist beliefs typically serve to deny and defend current social gender inequalities \cite{skewes_beyond_2018}. Gender essentialism also encompasses beliefs that gender is a binary (woman/man) and immutable, which is associated with increased prejudice towards trans and gender-diverse people \cite{jones2023gender}. Gender diverse and trans people are people who experience gender identities either outside of the gender binary of woman/man (e.g., non-binary) or gender identities not conventionally associated with their assigned sex at birth (e.g., assigned female at birth and identifies as a man) \cite{jones2023gender}. Gender essentialism is therefore also expressed as misogyny towards trans-women (people who were assigned male at birth and identify as women), both by perceiving trans-women as more masculine because of their biological sex \cite{gallagher2021gender}, or by hyper-sexualising trans-women while simultaneously viewing them as "not real women" \cite{robinson2023transamorous}.

\textbf{Toxic masculinity} refers to the way misogyny manifests in the strict enforcement of gender roles \cite{parent_social_2019}. This manifestation includes negative forms of masculinity, such as seeking to dominate and devalue others, particularly women, and homophobic language in cases where, for example, others are not perceived as masculine enough \cite{parent_social_2019}. Toxic masculinity can include using gendered or homophobic slurs in order to emasculate and attack other men. In addition, misogynist attacks directed at men often rely on insulting the women to whom the attacked men are related, such as a mother or a spouse \cite{Jane_2017}. 

\textbf{Gendered racism} captures the notion that women of colour face sexism and misogyny in more extreme forms than their white counterparts. For example, while it is true that women in general are over-sexualised and objectified \cite{ward_media_2016}, women of colour are often sexualised to an even greater degree \cite{wane_ruptures_2013}. We wanted our guidelines to be sensitive to the intersectional identities of women of colour and to acknowledge that attacks are often directed at them on the basis of both their gender and race. For instance, Black women are often criticised and discredited through the stereotype of the "angry Black woman" \cite{motro2022race} and East Asian women are stereotyped as being submissive and non-dominant more compared to white women \cite{berdahl2012prescriptive}.

\textbf{Post-feminism} is the view that gender inequalities are non-existent or minimal and women and men are currently fully equal \cite{jordan_conceptualizing_2016}. Post-feminist argumentation presents inequality as a thing of the past and denies current gender inequalities. Consequently, women can struggle to frame and communicate about their perceived experiences of gender inequality  \cite{morrison2005stop}. 

\textbf{Backlash} asserts that significant gender inequalities do exist, but against men rather than women \cite{jordan_conceptualizing_2016}. In some instances, such assertions are followed by the statements characterising gender equality as damaging to society \cite{jordan_conceptualizing_2016}. Backlash rhetoric is evident in statements claiming that feminism has taken away from men. This is a more recent form of misogyny which sustains the gender hierarchy with men at the top by denying women's inequality and also portraying measures and steps to advancing women in society as unnecessary and even as oppressive to men \cite{zehnter2021belief}. 

Post-feminism and backlash are not mutually exclusive, and movements often employ dual narratives that combine the two \cite{jordan_conceptualizing_2016}. To our reading, post-feminism and backlash are the most ignored manifestations of misogyny. An often ignored form of strong anti-feminist rhetoric is the idea of "misandry", the idea that men are now oppressed because of their gender, often as a result of feminism. We encourage a broadening of scope in the misogyny detection literature to acknowledge that such forms of rhetoric are not only inaccurate but have been and continue to be used to perpetuate stereotypes of feminists as "hating men" and to vilify the feminist movement and women more generally \cite{hopkins2024misandry}. 

We want to highlight that our guidelines do not in any way deny issues which may affect men. It is only when such issues are blamed on women and women's progress (which is often what terms like "misandry" are employed to do) that they are to be marked as misogynistic. For example, it is a fact that men experience higher suicide rates compared to women \cite{sher2020suicide}. This fact in isolation is not misogynistic, nor is exploring reasons for why it is true. However, blaming this fact on women's progress is a disservice to both women and men (i.e., at the very least it ignores the ways that men's adherence to traditional masculinity is linked to mental health outcomes and reluctance to seek help \cite{galvez2024exploring}).

\textbf{Internalised misogyny} is the idea that  patriarchal notions and standards can become ingrained in women \cite{szymanski_internalized_2009}. As a consequence, the devaluation of women through misogyny is perpetuated not only by men but also by women who themselves might reinforce these standards \cite{szymanski_internalized_2009}. For example, a woman might have internalised the belief that "men are more rational than women". Women with internalized misogyny are more likely to experience psychological distress related to sexist events \cite{szymanski_internalized_2009} and to show bias in favour of men over women \cite{han_using_2023, piggott_double_2004}. Annotators should be aware that even if they infer the speaker is a woman, it does not automatically mean their message is not misogynistic and what they have said should be evaluated in terms of the specified guidelines. It is also important for annotators who identify as women to be aware that they also, to some extent, may have internalised such beliefs and to engage in self-reflection during the annotation process.

Table \ref{tab:guidelines} provides some examples of each of these misogynistic phenomena. These points are developed further in our guidelines document, with supporting evidence and examples. We hope that these guidelines might help to start the process of developing a more standardised and less subjective guideline scheme for annotating corpora for instances of misogyny. Furthermore, it enables annotators to read more on certain topics if it might be helpful. Lastly, our guidelines contain a short section on Reddit jargon which annotators may be unfamiliar with. 

\section{Data Collection}

We used the PRAW (Python Reddit API Wrapper) package to collect data from 4 subreddits on Reddit. The subreddits are (1) r/MensRights, (2) r/Feminism, (3) r/LeftWingMaleAdvocates, (4) r/againstmensrights. 

The inclusion of the first 2 subreddits is motivated by \citet{rosen_berts_2023} who observed that posts from r/MensRights have significantly more similar rhetoric with posts from r/Feminism compared to other men's groups on Reddit. Men's Rights Activists (MRAs) frequently discuss women, feminism, and rights from an anti-feminist perspective \citep{kottig_at_2017}. When nuances in speech are not taken into consideration MRA rhetoric could easily be confused with feminism. Therefore, r/MensRights discourse provides an excellent data source for studying nuanced hate speech, and r/Feminism is a reasonable choice for a control group. 

Although the latter 2 subreddits are not explicitly discussed in the literature, we included them in our study based on our assessment that they would be suitable control groups for the first 2 subreddits. The subreddit r/LeftWingMaleAdvocates describes in its mission statement "We oppose right-wing exploitation of men's issues as a wedge to recruit men to inegalitarian traditional values. But we also oppose feminist attempts to deny male issues, or shoehorn them into a biased ideology that blames "male privilege" and guilt-trips men." \cite{reddit_rleftwingmaleadvocates_nodate}. 
The subreddit's mission statement emphasizes their commitment to discussing men's rights topics with an emphasis on left-wing politics. The statement suggests engagment with MRA activism while maintaining a deliberate differentiation from groups associated with the alt-right. The data sourced from r/LeftWingMaleAdvocates offer control for perceived political orientation for the data obtained from r/MensRights. The subreddit, r/againstmensrights self-desribes as "You might have heard that /r/MensRights is a moderate MRA hub.[...]Underneath the face lie toxic misogyny, GSMphobia, racism, and worse. The subreddit, and the movement itself, runs on hatred. We are here to expose the hatred and bring it to light." \cite{reddit_ragainstmensrights_nodate}. As the r/againstmensrights subreddit articulates opposition to r/MensRights both in its name and self-description, we considered it to be a suitable choice for a control group for perceived political orientation for the data obtained from r/MensRights (which holds an explicit alt-right stance). Additionally, we anticipated that this group would provide a valuable data source by discussing the same topics as r/MensRights but from a critical perspective. 

\section{Data Annotation}

Our annotators were two students volunteer research assistants (1 MSc Student and 1 BSc student) and one lab assistant (with an MSc). Annotators met as a group with one of the researchers to go over the guideline scheme in depth, address any questions and label a few test texts and discuss appropriate labels. In this meeting the researcher discussed with annotators the details regarding amounts of texts to be labelled, how much workload the annotators were willing to take and soft deadlines for the data annotations.

After the meeting, each annotator completed 100 annotations each week (~3 hours), where half the posts came from the control groups and half from the misogynistic group on Reddit. To promote objectivity, annotators were blind to which subreddit a post was taken from.

The annotation process lasted 4 weeks. Each week, the three annotators labelled 100 posts each, and together they labelled 150 posts in total. The allocation of posts was such that annotator A shared 50\% of their posts with annotator B and the remaining 50\% of their posts with annotator C. The same applied for annotators B and C. This ensured that 150 posts were annotated in total each week and that each post received 2 ratings, but that each annotator spent less time interacting with this very challenging data. 

Each week, annotators had a group meeting where the researcher checked in with them regarding their progress and mental state. The annotators shared advice with each other on how the managed the annotation workload and some strategies they found helpful for balancing the demands of the annotation work and their mental wellbeing. This advice is summarised in a section of our Guidelines document for future annotators. 

At the end of the 4 week annotation period, the researcher met with annotators for a final time to go over the disagreement texts. Together with the researcher, annotators revisited annotated texts and discussed their labels against the criteria to decide on which label was most appropriate.

\section{Annotation Results}

Inter-rate agreement was assessed with Cohen's kappa ($\kappa$). The first week of annotation during which the annotators were still training was excluded from this calculation. The reported $\kappa$ reflects the actual rating period. Moreover, though we had 3 raters, each post was rated only by 2 raters, so the ($\kappa$) value reflects inter-rater agreement between 2 raters for each item. The agreement between raters was substantial $\kappa=0.68$, which is statistically significant ($z=13, p < 0.001$). This indicates the raters agreed substantially on a level that was beyond random chance.  

\section{Case Study}
\subsection{Design}
In order to test how our guidelines perform when applied by LLMs without fine-tuning we employed a 3x3 design. We tested the performance of 3 different guideline conditions: (1) our guidelines, 2) Guest et al.'s guidelines and 3) no guidelines (control), against 3 different datasets: 1) our annotated dataset, (2) Guest et al.'s annotated dataset and (3) a curated dataset based on items from the psychometric scale of the Ambivalent Sexism inventory.

\section{Materials}

\begin{table*}[]
\caption{Summary of the 3 dataset names, content and their sources.}
\renewcommand{\arraystretch}{1.3}
\centering
\begin{tabular}{lll}
\toprule
Dataset Name  & Contents                                                                 & Source                                      \\
\midrule
Guest dataset & Annotated Reddit comments classified & \cite{guest_expert_2021} \\
& as misogynistic or non-misogynistic &\\
Our dataset   & Annotated Reddit posts classified  & Current paper \\
&   as misogynistic or non-misogynistic &  \\
ASI dataset &
  Misogynistic (hostile or benevolent) &
  Adapted from the Ambivalent Sexism Inventory \cite{glick_ambivalent_1996}\\
  & and non-misogynistic texts.&in the current paper \\
\bottomrule
\end{tabular}
\end{table*}

\subsection{Datasets}

\subsubsection{Choosing a comparable annotated dataset of misogyny} The dataset by Guest et al. \cite{guest_expert_2021} was selected for two reasons. Firstly, it contains annotated Reddit comments which are comparable to the Reddit posts in our annotated dataset. Secondly, datasets containing posts from X (formerly Twitter) only contain the post ID and require recollection of the texts via X's paid API. We wanted the datasets used and produced by this work to be openly accessible to the scientific community. 

Guest et al.'s dataset \cite{guest_expert_2021} contains in total 6534 annotated Reddit comments from a number of subreddits, including some from the men's rights subreddit. Despite being much larger than our dataset, only about 10\% of the data is labelled as misogynistic (N=699) compared to 43\% in ours (N=224). We test the LLMs on the full datasets, as we only prompt them and do not fine-tune them.

\subsubsection{Creating an Ambivalent Sexism testing dataset}
To create a dataset of Ambivalent Sexism items, we adapted the items from the Ambivalent Sexism inventory \cite{glick_ambivalent_1996} by keeping the original 22 items, as well as paraphrasing them to produce more statements with similar meanings. For example, the hostile item "Women are too easily offended" was rephrased to "Women get offended over anything", amongst others. Because these items are based on the psychological inventory, we did not need to annotate the new statements. Instead, items for which agreement would have indicated higher sexism were labels as "misogynistic", whereas items where lower agreement indicated sexism in the original inventory were labelled as "non-misogynistic".

We also rephrased items to have opposite meanings to create either misogynistic items from non-misogynistic items or vice-versa. For instance, the item "Feminists are making entirely reasonable demands of men." is reverse scored, so was labelled "non-misogynistic". Our re-phrasing reversed its meaning to "Feminists are making entirely unreasonable demands of men." which was labelled "misogynistic". 

This process yielded a dataset of 93 items in total (Benevolent: 36: , Hostile: 36, Non-misogynistic: 21), for testing the fidelity of the different guideline conditions to a well-known psychological theory of sexism. The resultant ASI dataset can be found at  \href{https://osf.io/3c8kp/overview?view_only=5ba9d2a050d34fc3bc84fdd27ef46e58}{OSF}.

\subsection{Selecting LLMs} Our design requires the prompt passed to the LLMs to contain both the guidelines and the annotated texts, some of which were quite lengthy. Therefore, our selection criteria for the 3 LLMs included: (1) the LLMs have large enough token context to fit both guidelines and comments/posts together in a singular prompt, (2) that these models are openly accessible for research purposes and (3) that they can be run on two H100 GPUs. The criteria narrowed this down to the following 3 models: Lamma 3.3 with 70 billion parameters\href{https://huggingface.co/meta-llama/Llama-3.3-70B-Instruct}{(meta-llama/Llama-3.3-70B-Instruct)} \cite{githubLlamamodelsmodelsllama3_2Main}, Mistral Large \href{https://huggingface.co/mistralai/Mistral-Large-Instruct-2411}{(mistralai/Mistral-Large-Instruct-2411)} \cite{mistral_large}, and Qwen2 with 72 billion parameters \href{https://huggingface.co/Qwen/Qwen2-72B-Instruct}{(Qwen/Qwen2-72B-Instruct)} \cite{qwen2}. All 3 models have a large context window of 128,000 tokens. 

\subsection{Coding schemes}

\subsubsection{A comparison of our guidelines to the Guest et al. guidelines}

\begin{table*}[h!]
   \caption{Summary of Guest et al. \cite{guest_expert_2021} guidelines taxonomy of misogyny. }
    \centering
    \begin{tabular}{cc}
        \toprule
      Taxonomy & Concepts \\ 
        \midrule
        misogynistic pejoratives & gendered slurs\\ \\
       misogynistic treatment & threatening language, physical violence, sexual violence, privacy\\
        & controlling, manipulation, seduction and conquest, other \\ \\
        
        misogynistic derogation &  intellectual inferiority, moral inferiority, sexual and/or physical limitations, other \\ \\

     gendered personal attacks & attacks made to individual women specifically referring their gender \\

    \bottomrule
    \end{tabular}
    \label{tab:guest_guidelines_summary}
\end{table*}

There are four important differences between the \cite{guest_expert_2021} guidelines and ours. First, our coding scheme defines sexism more broadly. \cite{guest_expert_2021} divide texts initially in a binary manner: misogynistic vs non-misogynistic, and within the misogynistic category they subcategorise mysogynistic texts as containing: (1) misogynistic pejoratives, (2) misogynistic treatment, (3) misogynistic derogation and (4) gendered personal attacks. These are then further divided to a number of related concepts, summarised in Table \ref{tab:guest_guidelines_summary}. Overall, when compared to our guidelines, the themes and examples covered by \cite{guest_expert_2021} fall under what we described as hostile sexism. Benevolent forms of sexism are not explicitly discussed in their coding scheme, as they are in ours.

Second,  our guideline schemes differ in their approach to language targetting feminist groups. The Guest et al. coding scheme does mention that categories like feminists are also gendered identities which are often targets of misogynistic language online \cite{guest_expert_2021}. However, the concept of anti-feminist backlash is not discussed or featured as a specific criterion for annotators. 

Third, while under person-directed abuse \cite{guest_expert_2021} do mention intersectional identities (e.g., black women), this topic is a brief note with one text example. Annotators are asked to identify "subgroup" identities, without much guidance as to the type of rhetoric used to attack women with intersectional identities. By contrast, in our guidelines we attempt to give a more broad overview of different intersectional identities women have (e.g. race, transgender) and how these identities might be targeted in particular ways (e.g. black women are stereotyped as being aggressive). 

Lastly, homophobia and other emasculating language that does not directly use a slur directed at women specifically is not considered a form of misogynistic language \cite{guest_expert_2021}. In our guidelines both of these types of language fall under concepts of toxic masculinity and possibly gender essentialism.

\subsection{Formatting the guidelines for prompting the LLMs}
We reviewed both original guideline documents so as to ensure the instructions provided to human annotators could feasibly be executed by the LLMs. In the case of Guest et al.'s guidelines some the instructions asked the human annotators to open links in comments and view the linked page when classifying. In the current case, the selected LLMs cannot open URLs so to maintain consistency between both guidelines and ensure the LLMs were given the same demands in the classification task we removed that instructions from the Guest et al. guidelines. 

Coding schemes are often several pages long as they need to be detailed and precise in order to guide annotators to be as accurate as possible in their work. This property poses a challenge when prompting LLMs, as the guidelines alone can consume a large part of the LLMs' context length. To address this problem we used the last open access Huggingface version of GPT \href{https://huggingface.co/openai/gpt-oss-20b}{(openai/gpt-oss-20b)} \cite{openai2025gptoss120bgptoss20bmodel} to summarise our guidelines and Guest et al.'s guidelines \cite{guest_expert_2021}. We chose a different LLM for the summarisation process to avoid introducing any bias in our design by having one of the main 3 LLMs summarise the guidelines and also follow those guidelines. The guidelines were given to GPT with the simple prompt to summarise them as clearly and concisely as possible. We set the max number of new tokens to 600 to restrict the model within a small but sufficient amount of words to adequatly summarise the concepts captured by both guidelines. Both summaries of the guidelines were read over by the researchers to ensure they were clear, concise and true to the original documents.

\subsection{LLM classification prompting} To prompt the LLMs to classify the items across all 3 datasets we used the same prompting template. The full prompt template is given in the colourbox below. The variable \textit{CODING\_SCHEME} was either our summarised coding scheme or the summarised coding scheme of Guest et al.\cite{guest_expert_2021} and the variable \textit{TEXT\_TO\_CLASSIFY} was occupied by each item in the dataset.

\begin{tcolorbox}[colback=black!5!white,colframe=black!75!white,title=Coding Schemes Prompt Template]
  Instruction: You are a strict classifier. \\
  - Never explain. \\
  - Never repeat the text or coding scheme. \\
  - Always follow this coding scheme strictly: \\  \textit{CODING\_SCHEME} \\
  - Only output JSON in this exact format: {"label": "<LABEL>"}.\\
  - <LABEL> must be \textbf{exactly one of these two values}: \\
  "misogynistic" or "non-misogynistic" \\
  - End output immediately after the closing brace. \\
  Input:\\
  \textit{TEXT\_TO\_CLASSIFY} \\
  Response:
  \tcblower
  Variables in capitalised \textit{ITALICS} are the arguments which changes across conditions.
\end{tcolorbox}

For the no coding scheme condition we removed the instructions about following the coding scheme as it was no longer relevant. You can read the exact prompt in the second colourbox below.  

\begin{tcolorbox}[colback=black!5!white,colframe=black!75!white,title= \textbf{No} Coding Scheme Prompt Template]
Instruction: You are a strict classifier. \\
- Never explain. \\ 
- Never repeat the text. \\
- Only output JSON in this exact format: {"label": "<LABEL>"}. \\
- <LABEL> must be **exactly one of these two values**: "misogynistic" or "non-misogynistic" \\
- End output immediately after the closing brace.\\
Input: \\
\textit{TEXT\_TO\_CLASSIFY} \\ 
Response: 
\tcblower
  Variables in capitalised \textit{ITALICS} are the arguments which changes across conditions.
\end{tcolorbox}

After prompting the LLMs with each text, memory was cleared to prevent memory problems and to ensure the response to the second item was not influenced by the first. During inference we disabled weight updating, meaning the models were not being fine-tuned.

\subsection{Output cleaning}: 

Though the prompt instructions specifically mentioned not to repeat the instructions, coding scheme or text itself and avoiding explanation, the LLMs would sometimes still produce more than just the JSON formatted label. To clean the data in a consistent and fair manner we used regex to find and keep only the first correctly formatted label provided in the LLM output for each text.

\subsection{Results}

As some of the datasets had a large class imbalance, we report macro- precision \footnote{$Precision_{macro}=\frac{Precision_{c_0} + Precision_{c_1}}{2}$, where $Precision_{c_n}= \frac{TP_{c_n}}{TP_{c_n} +FP_{c_n}} $ \\}, recall \footnote{ $ Recall_{macro} =\frac{Recall_{c_0}+ Recall_{c_1}}{2}$, where $Recall_c= \frac{TP_{c_n}}{TP_{c_n} +FN_{c_n}}$\\} and F1 score \footnote{$F1_{macro} = \frac{Precision_{macro} \times Recall_{macro}}{Precision_{macro} + Recall_{macro}} $}, which averages model performance across the two classes to consider both equally in the final metric. 

In Table \ref{tab:results1} we summarise the results of applying the same guidelines the dataset was annotated with compared to no guidelines. We see that across models supplying the guidelines in the prompt almost universally improves the performance of the LLMs on both the Guest et al. dataset and our dataset, relative to the condition where no guidelines are supplied. The only exception is Llama 3.3, for which when we provide our guidelines in the prompt there is an improvement in Precision, but a degraded performance in Recall and therefore an overall decrease in F1 score. Overall, there is strong evidence that the models are utilising the guidelines supplied via the prompt instructions and they achieve improved classification results. Interestingly, though the F1 score for the models for the Guest et al. dataset ranges from 0.82 at its lowest to 0.94 at the highest, the F1 score range for our dataset is much smaller, ranging from 0.40 to 0.65. This indicates that although 2 of 3 models improve when the guideline scheme is provided in the prompt, they still struggle to mimic the human annotator's labels as accurately as they did with the Guest et al data.

\begin{table*}[h!]
\caption{Model performance on the human annotated datasets with and without guidelines. Model performance is evaluated against human annotator labels. Reported estimates are rounded to 2 significant figures. We bold values to compare each model to itself across conditions (no guidelines vs with guidelines.)}
    \centering
    \hspace*{-0.3cm}
    \begin{tabular}{l|ccccc|ccccc}
    \toprule
    Model & Guidelines & Data & Precision & Recall & F1 score &Guidelines& Data & Precision& Recall & F1 score\\
    \midrule
        Llama 3.3 & No Guidelines & Guest Data 
        &0.83 & 0.71  & 0.71 &
        No Guidelines& Our Data 
        & 0.54& 0.53 & 0.47 \\

        Mistral Large & No Guidelines & Guest Data 
        &0.91 & 0.89 & 0.90&
        No Guidelines& Our Data 
        & 0.54& 0.54& 0.53 \\
        
         Qwen2 & No Guidelines & Guest Data 
        &0.90 & 0.90 &0.90 &
        No Guidelines& Our Data 
        &0.53 & 0.53& 0.52 \\

        \midrule
    
       Llama 3.3  & Guest Guidelines & Guest Data &
       \textbf{0.90} & \textbf{0.90} & \textbf{0.90} &
       Our Guidelines& Our Data &
       \textbf{0.59 }& \textbf{0.59}& \textbf{0.59}\\
       
        Mistral Large &  Guest Guidelines & Guest Data &
       \textbf{ 0.94 }& \textbf{0.92} & \textbf{0.93}&
         Our Guidelines& Our Data& \textbf{0.65}& \textbf{0.63}& \textbf{0.60}\\
        
        Qwen2 & Guest Guidelines & Guest Data 
        &\textbf{ 0.92 }& \textbf{0.92} & \textbf{0.92}&
         Our Guidelines& Our Data& \textbf{0.58}& \textbf{0.58 }& \textbf{0.57}\\

        \bottomrule
    \end{tabular}
    \label{tab:results1}
\end{table*}

We report the model performance metrics when we apply the models on the human annotated data with the same or different guidelines in Table \ref{tab:results2}. Comparing our guidelines vs the Guest et al. guidelines on the Guest dataset, we see that when our guidelines are applied the Recall of Llama 3.3 and Qwen2 improves and that of Mistrak Large remains the same. However, for Precision and F1-score all 3 models remains slightly higher (minimum F1 difference = 0.02, maximum F1 difference = 0.07). However, when we apply the Guest et al guidelines to our human annotated dataset the models' performance drops across all three metrics (Precision, Recall and F1 score) compared to when the models are given our guidelines. The differences in F1 score between the two conditions are also larger (minimum F1 difference = 0.23, maximum F1 difference = 0.35). This drop is performance is striking, as it is lower even than the performance of the 3 models on our dataset in the without any guideline scheme. Taken together, these results suggest that our guidelines perform similarly to the Guest et al guidelines to capture the phenomena that their human annotators were coding, but that the Guest et al guidelines cannot capture the phenomena that our human annotators were coding. 

We present the results comparing all 3 guideline schemes on the curated Ambivalent Sexism dataset in Table \ref{tab:results3}. We observe that best performance for Mistral Large and Qwen2 occurs when our guidelines are provided in the prompt, whereas for Llama 3.3 the best overall performance occurs with no guidelines. 

To analyse the performance of different coding schemes on the misogynistic and non-misogynistic classes we fit 3 Bayesian Logistic  Mixed Effects Regression models (one for each dataset). In all 3  regression models the binary outcome of incorrect/correct (0,1) label for that text entry was predicted by the categorical variables of the guideline scheme (none, Guest , ours), the true class (non-misogynistic, misogynistic) and the model (Llama, Mistral, Qwen). The reference levels for the categorical predictors were as follows: none (guideline scheme), non-misogynistic (true class), and Llama (model). We included random intercepts for each text entry, to reflect that different texts have different baseline probabilities of being classified correctly independent of the predictors. Additionally we used Leave-One-Out validation to test for a possible interaction between the guideline and true class predictors. Across all 3 datasets we find strong evidence that the interaction model outperforms the additive model in out of sample prediction; (1) Guest et al. dataset: \(\Delta\)ELPD= 626.0 , SD= 34.6 , (2) our dataset: \(\Delta\)ELPD= 263.0, SD= 14.7, (3) ASI dataset: \(\Delta\)ELPD= 17.7 , SD= 2.5. We therefore report the results for the three interaction models below.

\subsubsection{Guest et al. dataset} The model main effect estimates and random effect estimates for the Guest et al. dataset analysis are summarised in Table \ref{tab:guest_results_table}.  In the Guest et al. dataset, we find that compared to Llama 3.3, both Mistral (OR = 1.90, 95\% CrI[1.79, 2.02]) and Qwen (OR = 4.89, 95\% CrI[4.57, 5.24]) have increased odds of correctly classifying the texts. Mistral had a 90\% increase and Qwen had a 389\% increase in the odds of correctly classifying the texts. Controlling for the effect of the models, in the no guideline scheme condition, we find that the models had a 423\% increase in the odds of correctly identifying misogynistic texts relative to the non-misogynistic texts (OR= 5.23, 95\% CrI[4.02, 6.79]). 

\begin{figure}
\caption{Model estimates of the probability of classifying a text in the Guest et al. dataset correctly as a function of the interaction between the guideline scheme condition and the text class (misogynistic vs non-misogynistic). Bars represent the 95\% Credible Intervals.}
    \centering
    \includegraphics[width=1\linewidth]{ 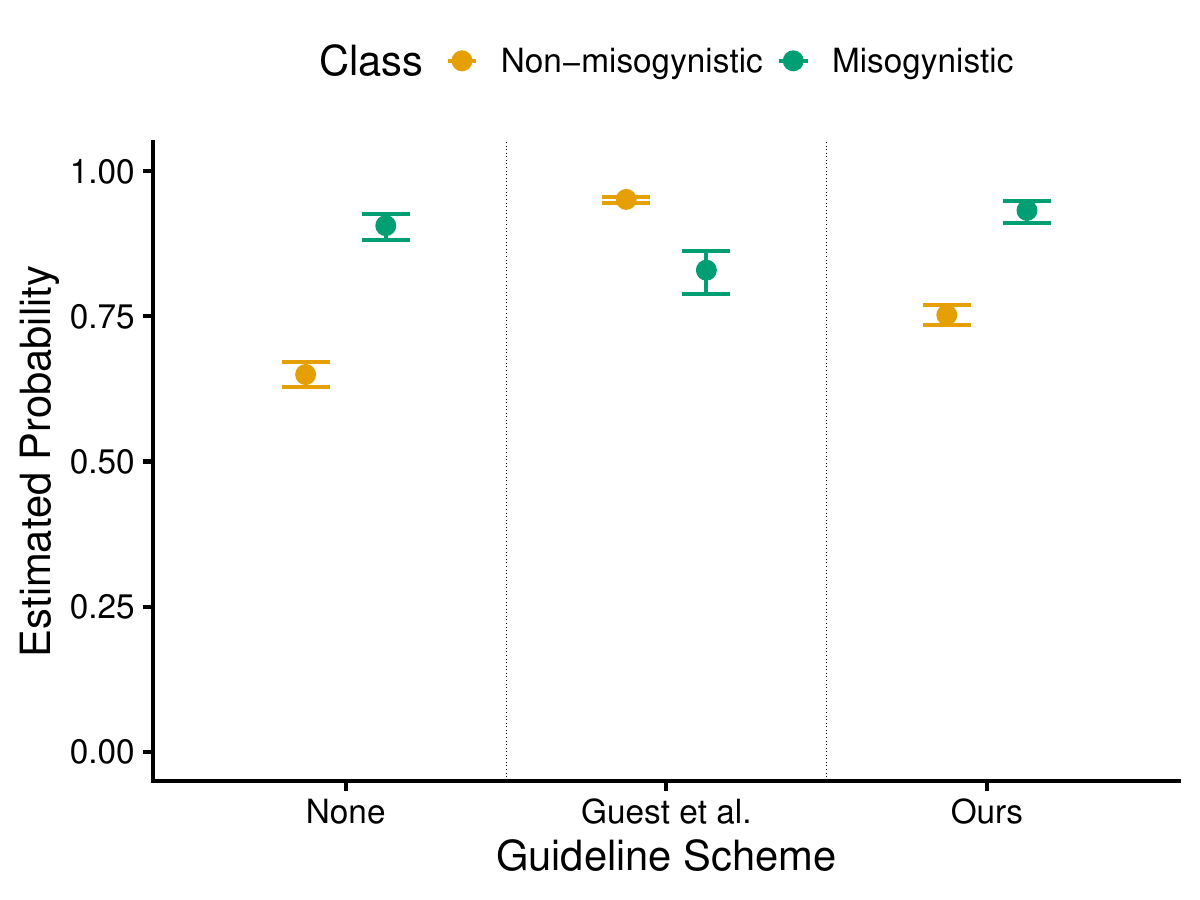}
    \label{fig:interaction_guestdata}
\end{figure}

Compared to the no guidelines condition, the Guest et al. guideline scheme had a 959\% increase in the odds of correctly identifying non-misogynistic texts (OR= 10.59, 95\% CrI[9.77, 11.53]). We also find that compared to the no guidelines condition, our guidelines had a 64\% increase in the odds of correctly classifying non-misogynistic texts (OR= 1.64, 95\% CrI[1.54 – 1.75]). However, relative to the no guideline scheme condition, the Guest et al. guidelines predicted a 95\% decrease in the odds of correctly classifying misogynistic texts (OR= 0.05, 95\% CrI [0.04, 0.06]). In contrast, our guidelines were not predicted to have an increase or decrease in the odds of correctly classifying misogynistic texts (OR= 0.87, 95\% CrI[0.70, 1.07]) when compared to the no guidelines scheme. Comparing our guidelines to the Guest et al. guidelines directly we observe a 85\% decrease in the odds of correctly classifying non-misogynistic posts (OR= 0.15, 95\% CrI[0.14, 1.17 ]). However, compared to the Guest et al. guideline scheme, our guideline scheme had a predicted 1736\% increase in the odds of correctly classifying misogynistic texts (OR = 18.36, 95\% CrI[14.59, 23.10 ]).  This interaction between coding schemes and text class are visualised in Figure \ref{fig:interaction_guestdata}.

\begin{table*}[h!]
\caption{Full results of the Bayesian Logistic Mixed Effects Regression fitted on the Guest et al. dataset. We report Odds Ratio estimates with 95\% credible intervals. Bold is used to mark estimates for which there was credible evidence of an association between the predictor and the outcome variable.}
    \hspace*{-0.3cm}
    \centering
    \begin{tabular}{ccc}
    \toprule
    Predictors &Odds Ratios	&  95\% Credible Intervals \\
    \midrule
    Intercept	&\textbf{1.86}	&\textbf{1.69 – 2.04}\\
    Codebook – Guest	&\textbf{10.59}	&\textbf{9.77 – 11.53}\\
    Codebook – Our &\textbf{1.64}	&\textbf{1.54 – 1.75}\\
    Class– misogynistic	&\textbf{5.23}	&\textbf{4.02 – 6.79}\\
    Model Name – Mistral Large	&\textbf{1.90}	&\textbf{1.79 – 2.02}\\
    Model Name  Qwen2 &	\textbf{4.89}&	\textbf{4.57 – 5.24}\\
    Codebook – Guest:Class – misogynistic&	\textbf{0.05}&	\textbf{0.04 – 0.06}\\
    Codebook – Ours : Class – misogynistic	&0.87	&0.70 – 1.07\\
    \midrule
    Observations &59065 &\\
    \bottomrule
    \end{tabular}
    \label{tab:guest_results_table}
\end{table*}

\subsubsection{Our dataset} The model main and random effect estimates are summarised in Table \ref{tab:ourdata_results_table}. We do not find that Mistral Large (OR=1.13, 95\% CrI[0.95, 1.35]) or Qwen (OR= 1.03, 95\% CrI[0.86, 1.23]) predicted the odds of classifying a text correctly. In the no guidelines condition we find that misogynistic texts have a decrease in the odds of being classified correctly relative to non-misogynistic posts (OR=0.69, 95\% CrI[0.50, 0.95]). In other words, in the no guidelines conditions the LLMs misclassified misogynistic texts (FN) more frequently than non-misogynistic texts. 

\begin{figure}
\caption{Model estimates of the probability of classifying a text in our dataset correctly as a function of the interaction between the guideline scheme condition and the text class (misogynistic vs non-misogynistic). Bars represent the 95\% Credible Intervals. We project the average prediction of the 3 LLMs for each text to show model fit to the data.}
    \centering
    \includegraphics[width=1.\linewidth]{ 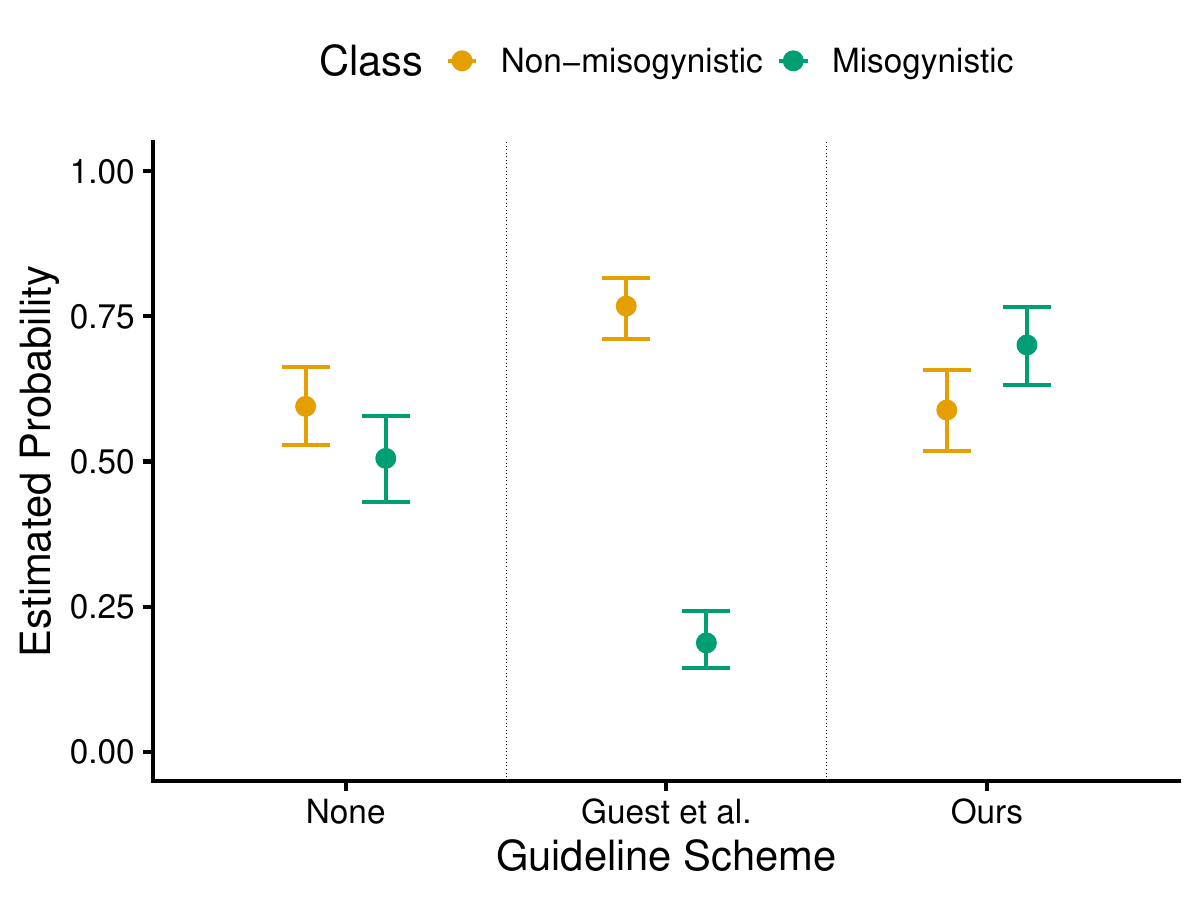}
    \label{fig:interaction_mydata}
\end{figure}

There was evidence of an interaction between the guideline scheme used and the class of the texts. Compared to the no guidelines condition, the Guest et al. guidelines predicted a 125\% increase in the odds of classifying non-misogynistic posts correctly (OR=2.25, 95\% CrI[1.83, 2.80]), but also a marked decrease in the odds of correctly classifying misogynistic texts (OR= 0.10, 95\% CrI[0.08, 0.13]). In other words, the Guest et al. guidelines resulted in more TN predictions but at the expense of more FN predictions. Compared to the no guidelines condition, our guidelines did not predict a decrease or increase in the odds of classifying the non-misogynistic texts correctly (OR= 0.98, 95\% CrI[0.80, 1.18]), but they predict a an overall 60\% increase in the odds of classifying misogynistic texts correctly (OR = 2.35, 95\% CrI[1.79, 3.11]). In other words, our guidelines resulted in more TP classifications, but not at the expense of FP classifications. After accounting for the interaction effect, in the misogynistic texts class, our guidelines are predicted to have a 2257\% increase in the odds of classifying misogynistic text correctly compared to the Guest et al. guidelines (OR= 23.57, 95\% CrI[16.61, 33.44]). The interaction is visualised in Figure \ref{fig:interaction_mydata}

\begin{table*}[h!]
 \caption{Full results of the Bayesian Logistic Mixed Effects Regression fitted on our dataset. We report Odds Ratio estimates with 95\% credible intervals. Bold is used to mark estimates for which there was credible evidence of an association between the predictor and the outcome variable.}
    \centering
    \begin{tabular}{ccc}
    \toprule
    Predictors	& Odds Ratios & 95\% Credible Intervals \\
    \midrule
    Intercept	&\textbf{1.47}	&\textbf{1.12 – 1.97}\\
    Codebook – Guest	&\textbf{2.25}	&\textbf{1.83 – 2.80}\\
    Codebook – Ours &	0.98	&0.80 – 1.18\\
    Class – misogynistic & \textbf{0.69}	& \textbf{0.50 – 0.95}\\
    Model Name– Mistral Large &1.13	 & 0.95 – 1.35 \\
    Model Name – Qwen2	& 1.03	& 0.86 – 1.23 \\
    Codebook – Guest: Class – misogynistic &	\textbf{0.10}	&\textbf{0.08 – 0.13}\\
    Codebook – Ours: Class – misogynistic	& \textbf{2.35}	&\textbf{1.79 – 3.11}\\
    \midrule
    Observations &	4653 & \\
    \bottomrule
    \end{tabular}
    \label{tab:ourdata_results_table}
\end{table*}

\subsubsection{ASI dataset} We do not find that Mistral Large (OR= 1.23, 95\% CrI[0.86, 1.74]) had credibly different odds of classifying texts correctly compared to LLama 3.3. We do find that Qwen had a 30\% decrease in the odds of classifying texts correctly relative to Llama 3.3 (OR= 0.70, 95\% CrI[0.50,0.97]). Controlling for predicted effect of the models, we find that in the no guideline scheme condition there was an increase or decrease in the odds odds of correctly classifying misogynistic texts relative to the non-misogynistic texts (OR= 1.28, 95\% CrI[0.82 – 2.01]). Compared to the no guideline scheme condition, the Guest et al. guidelines were predicted to have 35\% decrease in the odds of correctly classifying non-misogynistic texts (OR= 0.65, 95\% CrI[0.44 – 0.94]). Our guidelines did not show a marked decrease or increase in the odds of correctly classifying non-misogynistic texts relative to the no guideline scheme condition (OR = 1.00, 95\% CrI[0.69, 1.45]). Compared to the no guideline scheme condition, the Guest et al. guidelines predicted a 56\% decrease in the odds of correctly classifying misogynistic texts (OR = 0.44 , 95\% CrI[0.30, 0.67]). In contrast, our guidelines were predicted to have a 67\% increase in the odds of correctly classifying the misogynistic texts, relative to the no guidelines condition (OR= 1.67, 95\% CrI[1.13, 2.46]).

\begin{figure}
\caption{Model estimates of the probability of classifying a text in the ASI dataset correctly as a function of the interaction between the guideline scheme condition and the text class (misogynistic vs non-misogynistic). Bars represent the 95\% Credible Intervals.}
    \centering
    \includegraphics[width=1\linewidth]{ 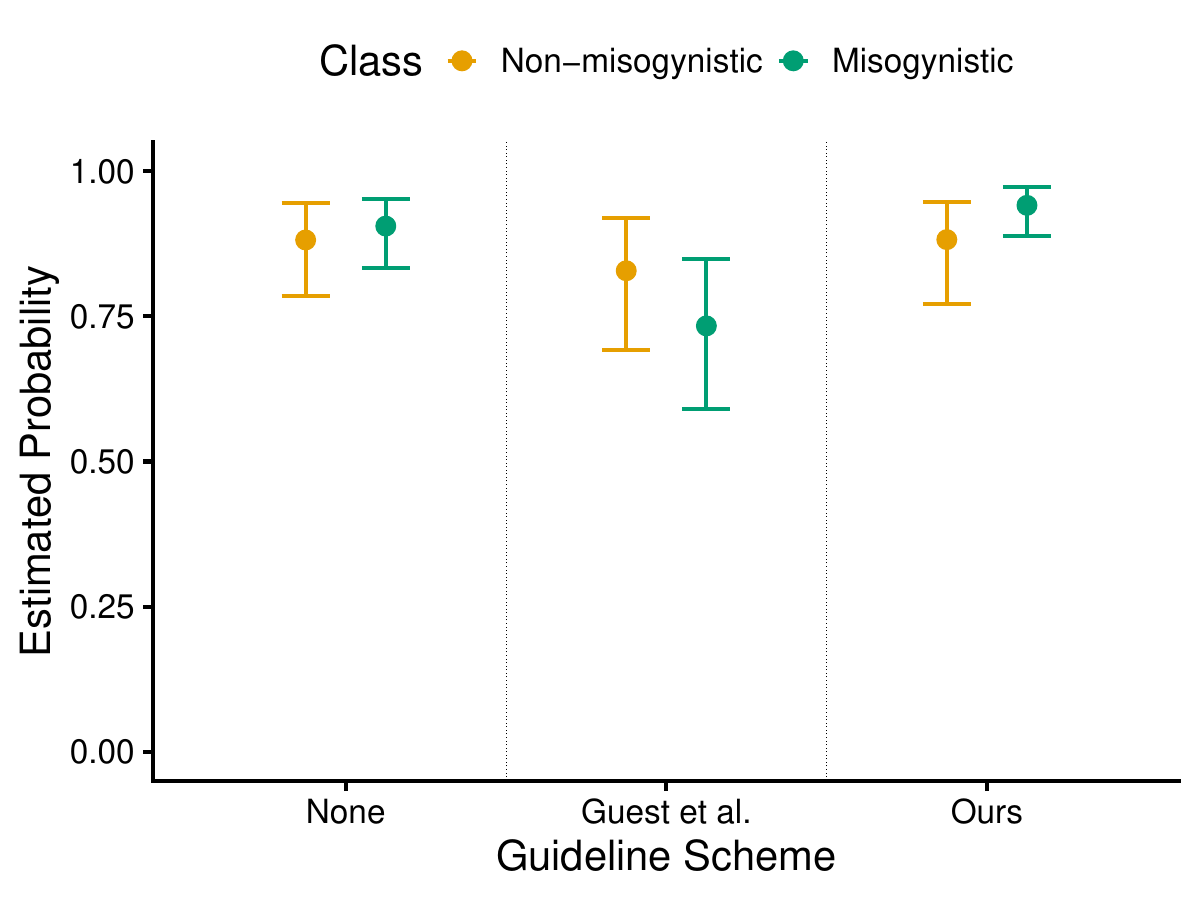}
    \label{fig:interaction_ASI}
\end{figure}

Comparing the Guest et al guideline to our guideline directly, there is no predicted difference between them in the odds of classifying non-misogynistic texts correctly (OR= 1.55, CrI[0.95, 2.50]. However, compared to the Guest et al. guidelines, our guideline scheme had a 274\% increase in the odds of correctly classifying misogynistic texts correctly (OR = 3.74, 95\% CrI[2.23, 6.49]). This interaction being the guideline schemes and the text class is visualised in Figure \ref{fig:interaction_ASI}

\begin{table*}[h!]
 \caption{Full results of the Bayesian Logistic Mixed Effects Regression fitted on the ASI dataset. We report Odds Ratio estimates with 95\% credible intervals. Bold is used to mark estimates for which there was credible evidence of an association between the predictor and the outcome variable.}
    \centering
    \begin{tabular}{ccc}
    \toprule
    Predictors & Odds Ratios	& 95\% Credible Intervals \\
    \midrule
    Intercept	&\textbf{7.46}	&\textbf{3.65 – 17.18} \\
    Codebook – Guest	&\textbf{0.65}	&\textbf{0.44 – 0.94}\\
    Codebook – Ours &	1.00	&0.69 – 1.45 \\
    Class – Misogynistic	& 1.28	&0.82 – 2.01 \\
    Model Name – Mistral Large	& 1.23	& 0.86 – 1.74\\
    Model Name – Qwen2 &	\textbf{0.70}	& \textbf{0.50 – 0.97 }\\
    Codebook – Guest: Class – misogynistic	&\textbf{0.44}	&\textbf{0.30 – 0.67}\\
    Codebook – Ours: Class – misogynistic	&\textbf{1.67}	&\textbf{1.13 – 2.46} \\
    \midrule
    Observations	&837 & \\
    \bottomrule
    \end{tabular}
    \label{tab:placeholder}
\end{table*}

\begin{table*}[h!]
\caption{Model performance on the human annotated datasets comparing between using with the same guidelines and different guidelines to those provided to human annotators. Model performance is evaluated against human annotator labels. Reported estimates are rounded to 2 significant figures. We bold values to compare each model to itself across conditions (same guidelines vs different guidelines.)}
    \centering
    \hspace*{-0.5cm}
    \begin{tabular}{l|ccccc|ccccc}
    \toprule
    Model & Guidelines & Data & Precision & Recall & F1 score &Guidelines& Data & Precision& Recall & F1 score\\
    \midrule
    
    Llama 3.3  & Guest Guidelines & Guest Data &
       \textbf{0.90} & \textbf{0.90} & \textbf{0.90} &
  Our Guidelines& Guest Data &
   0.87 & 0.85 & 0.85 \\
  
Mistral Large &  Guest Guidelines & Guest Data &
       \textbf{ 0.94 }& \textbf{0.92} & \textbf{0.93}&
  Our Guidelines& Guest Data &
  0.88& 0.81 &0.82 \\

  Qwen2 & Guest Guidelines & Guest Data 
        &\textbf{ 0.92 }& \textbf{0.92} & \textbf{0.92}&
  Our Guidelines& Guest Data &
  0.92& 0.89 &0.90 \\

  \midrule

    Llama 3.3 & Guest Guidelines & Our Data &
    0.48& 0.49  & 0.46  
&Our Guidelines& Our Data &
       \textbf{0.59 }& \textbf{0.59}& \textbf{0.59}\\
  
Mistral Large & Guest Guidelines & Our Data
&0.49&0.49  &0.47
  &Our Guidelines& Our Data& \textbf{0.65}& \textbf{0.63}& \textbf{0.60}\\

Qwen2 & Guest Guidelines & Our Data
& 0.46& 0.47 &  0.44
  & Our Guidelines& Our Data& \textbf{0.58}& \textbf{0.58 }& \textbf{0.57}\\

\bottomrule
    
    \end{tabular}
    \label{tab:results2}
\end{table*}

\begin{table}[h!]
\caption{Model performance on the curated Ambivalent Sexism Inventory dataset comparing the 3 guideline conditions. Model performance is evaluated against the inventory-informed labels. Reported estimates are rounded to 2 significant figures. We bold values to compare how models perform with different guidelines (no guidelines, Guest et al. guidelines and our guidelines.)}
    \centering
     \hspace*{-0.3cm}
    \begin{tabular}{l|cccc}
    \toprule
    Model & Guidelines & Precision & Recall & F1 score \\
    \midrule
    
    Llama 3.3 & No Guidelines &\textbf{0.89} &\textbf{ 0.73} & \textbf{0.77} \\

    Mistral Large & No Guidelines & 0.67 & 0.70 & 0.68 \\
    
    Qwen2 &  No Guidelines & 0.65 & 0.69 & 0.65 \\

    \midrule

    Llama 3.3 & Guest Guidelines & 0.53 & 0.54 & 0.52 \\

    Mistral Large &Guest Guidelines & 0.61 & 0.65 & 0.61\\

     Qwen2 & Guest Guidelines & 0.62 & 0.66 & 0.55 \\

     \midrule

      Llama 3.3 & Our Guidelines & 0.63 & 0.60 & 0.61 \\
      
 Mistral Large &  Our Guidelines & \textbf{0.93}& \textbf{0.71 }& \textbf{0.76} \\
 
  Qwen2 &  Our Guidelines 
  & \textbf{0.67} & \textbf{0.71} & \textbf{0.68} \\

\bottomrule
    
    \end{tabular}
    \label{tab:results3}
\end{table}

\section{Discussion}
We have outlined a new guideline scheme which is grounded in psychological theory and empirical work. Though the guideline scheme itself encompasses many concepts which may be challenging for annotators to apply, we have demonstrated that our annotators were able to apply the scheme with adequate levels of agreement (Cohen's \(\kappa\)=0.68). These levels of agreement are notably higher than what is typically reported in the literature for misogyny detection (e.g., \cite{guest_expert_2021, anzovino2018automatic}). Both the guidelines and annotated dataset are publicly available. 

The dataset itself is small compared to previously released datasets in the literature. However, new methods for fine-tuning LLMs with smaller datasets have shown promising results (e.g. \cite{hulora, dettmers2023qlora}. At the same time, both this coding scheme and dataset are meant to serve as a stepping stone for the research community working on misogyny detection, and we invite other researchers to use and expand on the materials presented here in their work, to collectively promote a more psychologically grounded and consistent approach to misogyny detection. 

Additionally, we have exemplified the importance of implementing theoretical and empirical knowledge in our guideline schemes and datasets for misogyny detection in our case study. Overall, across all 3 datasets we find strong evidence that our guideline scheme improves the classification of misogynistic texts relative to the scheme by Guest et al. \cite{guest_expert_2021}. Additionally, we observe that the Guest et al. guideline scheme degrades the classification of misogynistic texts for both our dataset and the adapted ASI dataset. We attribute these findings to the differences in the content covered by our coding schemes. The Guest et al. guideline scheme overemphasises hostile expressions of misogyny compared to ours, which pays particular attention to benevolent sexism amongst less well known concepts in the misogyny detection literature. 

Overall, the LLMs underperform in the classification of our dataset and the ASI dataset relative to the Guest et. al dataset. We believe this gap in performance scores (macro- Precision, Recall and F1) points to the inbuilt biases LLMs have towards the mainstream view of misogyny, which more closely aligns with the  Guest et al. guideline scheme. There is ample prior work reporting a number of social biases, like gender stereotypes, are reflected in LLM outputs (including Llama and Mistral) \cite{kotek2023gender,chen2025causally}. Even seemingly more complex cognitive biases, such the association of personhood with men over women \cite{hamilton1991masculine}, have been shown to be reflected in word embeddings extracted from a number of corpora \cite{bailey2022based}. It should not be understated that the learned social biases in LLMs will pose a great challenge to their use for misogyny detection.

\subsection{Implications for the use of LLMs in the detection of misogyny}

LLMs are currently being used to detect \cite{dev2025beyond, morbidoni2023can, altin2025investigating} and taxonomise misogynistic language \cite{ailneni2025automatically}. The suitability of LLMs for the categorisation of misogyny in both cases is undetermined, given the social biases reflected in LLM outputs. Our results in particular contradict prior claims that clear definitions in LLM prompting can lead to the precise detection of misogyny \cite{dev2025beyond}. It is necessary to further examine the suitability of LLMs for misogyny detection.

It is the case that LLMs can rapidly update their 'knowledge' through exposure to new data. This property is in theory ideal for a problem like misogyny, where the language and rhetoric are fast evolving \cite{ribeiro2021evolution}. However, this might also prove to be a weakness . The exposure of LLMs to misinformation has been found to cause knowledge drift \cite{fastowski2024understanding}. By the same token, there is a risk that the more LLMs are exposed to new forms of misogynistic real world data, learned social biases are reinforced. 

One obvious solution is to try and mitigate these acquired social biases. Some studies report optimistic estimates. For example, fine-tuning LLMs can help reduce stereotypes up to 40\% in one epoch \cite{raj2024breaking}. However, currently there is a lot of uncertainty surrounding the efficacy of available methods for reducing social bias in LLMs. Both in-context learning and fine-tuning have a reportedly moderate reduction effect for biases \cite{liu2024confronting}. However,  when we consider the fairness metric gap between different groups it is still larger compared to machine learning models, like random forest and shallow neural networks \cite{liu2024confronting} and in some cases neural networks can outperform LLM prompting for the classification of misogyny \cite{altin2025investigating}, (at least for low resource languages). Though LLMs are undeniably powerful tools, currently their performance for misogyny detection is limited by inherited social biases and more work is needed to shed light on how to best mitigate those.

\section{Conclusion}
The current paper introduces a new guideline scheme and dataset for the annotation of misogynistic language, which are grounded in psychological theory and show improved inter-rater agreement relative to prior work. Though these guidelines could improve the classification of misogynistic language relative to the baseline and another existing set of guidelines, the overall performance of the LLMs was poor. The LLMs underperformed because of inbuilt biases, which reflect mainstream views of misogyny that contradict the psychological literature. Even when psychologically grounded guidelines are provided in the prompt the biases are not sufficiently mitigated. This work provides a new set of tools for the improvement of misogyny datasets and raises important implications for the application of LLMs for misogyny detection. 

\section{Limitations}

As previously mentioned the dataset we have annotated is small in size, when compared to other datasets in the literature. Whilst techniques for fine-tuning and training models on smaller datasets are being developed, we acknowledge the need for this dataset to be expanded. Ideally new sources of texts would be included in future work, to expand the communities and expressions of misogyny covered across the web.   

As with any work in hate speech, our current guidelines are limited to the current linguistic and cultural climate. Both linguistic and cultural shifts could mean that in the future our guidelines will need to be updated to reflect new formalised concepts of psychology in misogyny and likely new forms of hateful rhetoric. 

The current paper focuses only on binary classification. Further subcategorisation of misogynistic texts can add value (e.g. \cite{jha2017does}) to datasets of misogyny detection. The  work presented here is a first step towards theoretically grounding the guideline schemes and datasets for misogyny detection. Future work should expand the dataset by further labelling the particular ways misogyny is expressed in the texts. We do caution, however, that the concepts covered in the present guidelines are not mutually exclusive. This can pose challenges for future work aiming to categorise misogynistic texts into subcategories, as this will likely require multiple labels to be applied to each text. 

\section{Ethics}
This work received ethical approval by the School of Informatics Ethics Committee at the University of Edinburgh (Reference number:851495).

In terms of the data annotation, we were mindful that our annotators had to read and classify content which could impact their mental wellbeing if exposed to it for too long. Annotators were reminded at multiple stages that their work was voluntary and could be stopped at any point with no repercussions and this was also outlined in the guideline document. Additionally, after discussion with the annotators we decided a week was an be appropriate amount of time to annotate each set of posts. Annotators were encouraged to split the workload as their schedules permitted and their mental wellbeing demanded. Moreover, we held weekly check-in meetings with annotators to check their progress and state. At the end of the annotation period we also included a section to the guidelines for future annotators, which includes advice for future annotators based on their experience during this process.  

When undertaking the work outlined here we were mindful that we do not train existing closed LLMs and that all the work we detail here will be openly accessible. To this end we only used open access LLMs, which we could run on local GPUs and ensured that all our materials are all made public for the community. 

Additionally, we wanted to respect that in the future Reddit users may wish to deactivate their accounts or not be associated with their past posts, even if their accounts are anonymous. At the same time, many in the research community rely on the distributional information of which users generate misogynistic posts in a sample. To balance the rights of users and the needs of the scientific community we replaced usernames with randomly generated author IDs.

\begin{acks}

This work was supported by the UKRI Centre for Doctoral Training in Natural Language Processing, funded by UKRI (grant EP/S022481/1). We want to thank our volunteer annotators, whom without this work would not have been possible. 
\end{acks}

\section{Generative AI usage statement} We declare that Generative AI was not used in the writing of this paper.

\bibliographystyle{ACM-Reference-Format}
\bibliography{citations}




\end{document}